# Univariate Polynomial Equation Providing Models of Thermal Lattice Boltzmann Theory

Jae Wan Shim

*Interdisciplinary Fusion Technology Division, KIST & UST 136-791, Seoul, Korea*

A univariate polynomial equation is presented. It provides models of the thermal lattice Boltzmann equation. The models can be accurate up to any required level and can be applied to regular lattices, which allow efficient and accurate approximate solutions of the Boltzmann equation. We derive models satisfying the complete Galilean invariant and providing accuracy of the 4th-order moment and *beyond*. We simulate thermal shock tube problems to illustrate the accuracy of our models and to show the remarkably enhanced stability obtained by our models and our discretized equilibrium distributions.

The kinetic theory of gases constitute the statistical theory of the dynamics of mechanical systems based on a simplified molecular description of a gas, from which macroscopic physical properties of the gas can be derived by using a velocity distribution function which describes how molecular velocities are distributed on average. The Boltzmann equation describes how collisions and external forces cause the velocity distribution to change. Through the Chapman-Enskog expansion of the Boltzmann equation, we can obtain the Euler, Navier-Stokes, Burnett equations etc. according to the orders of approximation [1]. The lattice Boltzmann equation (LBE) is a discretized version of the Boltzmann equation in phase space and time [2,3]. Originally, the LBE was developed from the lattice-gas cellular automata [4-7], where fluids are simulated by using fictitious molecules hopping in regular lattices. The molecules collide with one another on the nodes of a lattice and move to other nodes successively. In the LBE, the fictitious molecules are replaced with a molecular probability distribution. There exist various models of the LBE according to the shape of lattice, the dimension of space, and the number of discrete velocities. Increasing the number of discrete velocities is a way to obtain models providing the complete Galilean invariant [8] and the correct thermal results for compressible flows. However, the ratios of discrete velocities should be rational for collisions to occur on the nodes of regular lattices; otherwise additional effort is required. It is proposed to use the preservation of the norm and the orthogonality of the Hermite polynomial tensors for obtaining models, however, this framework is difficult to obtain higher-order models because we should derive and calculate a system of equations for each model [9]. Using the minimization of an entropy function provides models applicable to regular lattices [10,11], however, it has the same problem with the aforementioned framework and is still limited to isothermal flows.

In this Letter, we derive a univariate polynomial equation whose variable is a discrete velocity. The coefficients of the equation are composed of the ratios between discrete velocities. Therefore, the problem to find a model of any required level of accuracy, applicable to regular lattices, is reduced to a problem only to solve the univariate polynomial equation. We also discuss the dependence of the stability of models on discretized equilibrium distributions and propose robust ones. We present explicitly several models to simulate the thermal shock tube problem. To illustrate the accuracy our models, we compare the simulation results with the analytical solution of the Riemann problem [12]. We also show the remarkable improvement of stability obtained by our models and our discretized equilibrium distributions.

The Boltzmann equation with the BGK collision term is $\partial_t f + \mathbf{V} \cdot \nabla f = -(f - f^{eq})/\tau$. The velocity distribution function $f$ is defined by the infinitesimal quantity $f\,d\mathbf{x}d\mathbf{V}$ which is the number of particles having velocity $\mathbf{V}$ in an infinitesimal element of phase space $d\mathbf{x}d\mathbf{V}$ at position $\mathbf{x}$ at time $t$. The relaxation time $\tau$ adjusts the tendency of $f$ to approach the Maxwell-Boltzmann (MB) distribution $f^{eq}$ due to collision. Macroscopic physical properties are obtained by $\rho\{1, \mathbf{U}, e\} = \int f\{1, \mathbf{V}, 2^{-1}\|\mathbf{V}-\mathbf{U}\|^2\} d\mathbf{V}$ where $\rho$ is number density, $\mathbf{U}$ macroscopic velocity, and $e$ energy per unit of mass. We can relate $e$ with temperature $T$ by $e = Dk_B T/(2m_g)$ where $D$ is the dimension of space, $k_B$ the Boltzmann constant, and $m_g$ molecular mass. The MB distribution is $f^{eq} = \rho(\pi\theta)^{-D/2}\exp\left(-\|\mathbf{v}-\mathbf{u}\|^2\right)$ where dimensionless variables are defined by $\theta \equiv 2k_B T/m_g$, $\mathbf{v} \equiv \theta^{-1/2}\mathbf{V}$, and $\mathbf{u} \equiv \theta^{-1/2}\mathbf{U}$. The discretized version of the Boltzmann equation in phase space and time can be written by $f_i(\mathbf{x}+\mathbf{V}_i, t+\Delta t) - f_i(\mathbf{x},t) = -[f_i(\mathbf{x},t) - f_i^{eq}(\mathbf{x},t)]/\tau$ where $f_i(\mathbf{x},t)$ is the probability for a particle to exist in a lattice site $\mathbf{x}$ at time $t$ with discrete velocity $\mathbf{V}_i$. The essential work of the discretization is to find the discretized MB distribution $f_i^{eq}$ satisfying

$$\int \mathbf{v}^m f^{eq}(\mathbf{v})d\mathbf{V} = \sum_i \mathbf{v}_i^m f_i^{eq}(\mathbf{v}_i) \qquad (1)$$

to conserve physical properties such as mass, momentum, pressure tensor, energy flux, and the change rate of the energy flux etc., which are obtained by $m$-th order moments of $\mathbf{V}$, i.e. $\int \mathbf{V}^m f^{eq}(\mathbf{V})d\mathbf{V}$. If we can express $f^{eq}$ by a series expansion $f^{eq}_E$ having the form of

$$f^{eq}(\mathbf{v}) \approx f^{eq}_E(\mathbf{v}) = \exp(-v^2)P^{(k)}(\mathbf{v}) \qquad (2)$$

where $P^{(k)}(\mathbf{v})$ is a polynomial of degree $k$ in $\mathbf{v}$ and $v^2 = \mathbf{v}\cdot\mathbf{v}$, we can find $f_i^{eq}$ in the form of

$$f_i^{eq}(\mathbf{v}_i) = w_i P^{(k)}(\mathbf{v}_i) \qquad (3)$$

where $w_i$ are constant coefficients. By applying (2) and (3) to (1), we obtain

$$\int \exp(-v^2) P^{(m+k)}(\mathbf{v})d\mathbf{v} = \sum_i w_i P^{(m+k)}(\mathbf{v}_i). \qquad (4)$$

We begin our systematic procedure with 1-dimensional space. Formula (4) is satisfied for any polynomial of degree $m+k$, if and only if

$$\int v^n \exp(-v^2)dv = \sum_i w_i v_i^n \qquad (5)$$

for any integer $n$ such that $0 \le n \le m+k$. We can calculate the left side of (5) as

$$\int v^n \exp(-v^2)\, dv = \begin{cases} \Gamma((n+1)/2) & \text{for } n = 2\alpha, \\ 0 & \text{for } n = 2\alpha+1 \end{cases} \qquad (6)$$

where $\alpha$ is any non-negative integer and $\Gamma$ is the Gaussian Gamma function which can be expressed by the double factorial as $\Gamma((n+1)/2) = \sqrt{\pi}(n-1)!!/2^{n/2}$.

The system of equations (5) can become more concise by considering symmetry. Let $M^1|q$ be the 1-dimensional $q$-velocities models such that their discrete velocities $v_i$ and weight coefficients $w_i$ are

$$\begin{aligned} & v_1 = 0,\ v_{2i} > 0,\ v_2 < v_4 < \ldots < v_{q-1}, \\ & v_{2i} = -v_{2i+1},\ \text{and}\ w_{2i} = w_{2i+1}\ \text{for}\ i = 1,2,\ldots,[q/2] \end{aligned} \qquad (7)$$

where $[x]$ is the greatest integer that is less than or equal to $x$. Note that we regard $q$ as an odd number to include the zero velocity $v_1$. To use regular lattices, the ratios of $v_i$ should be rational numbers, therefore, we have the constraints of

$$v_{2(i+1)}/v_2 = p_{2(i+1)}/p_2 = \bar{p}_{2(i+1)} \quad \text{for}\ i = 1,2,\ldots,[q/2]-1 \qquad (8)$$

where $p_2$ and $p_{2(i+1)}$ are relatively prime and $p_{2(i+1)} > p_{2i}$. These models have $2q$ variables composed of $v_i$ and $w_i$, but have only $q$ unknown variables by their symmetry (7). If $p_2$ and $p_{2(i+1)}$ are given, we can express all $v_i$ by $v_2$. Consequently, we have $n' \equiv (q+3)/2$ unknown variables. The variables defined by (7) satisfy $\Xi(n) \equiv \sum w_i v_i^n = 0$ for any odd number $n$. Therefore, to find $n'$ unknown variables, we need $n'$ equations and they are

$$\{\Xi(n) = \Gamma((n+1)/2) \mid n = 0, 2, \ldots, 2(n'-1)\}. \qquad (9)$$

If the solution of (9) exists, it satisfies the polynomial of degree $m+k$ up to $2(n'-1)+1$ ($=q+2$) in (4). Let us consider a $q'$-velocities model which does not possess the zero velocity. It satisfies the polynomial up to $m+k = q'+1$ where $q'$ is an even number. This means a $q$-velocities model satisfies the same order of the moment accuracy $m+k$ as a $(q+1)$-velocities model for an odd number $q$. Therefore, an odd number is preferred for the number of discrete velocities from the viewpoint of the minimization of discrete velocities.

The system of equations (9) can be reduced to a univariate polynomial equation. We define

$$\mathbf{w} = \begin{bmatrix} w_2 \\ w_4 \\ \vdots \\ w_{q-1} \end{bmatrix},\ \mathbf{A} = \begin{bmatrix} \bar{p}_2^{\,2} & \bar{p}_4^{\,2} & \cdots & \bar{p}_{q-1}^{\,2} \\ \bar{p}_2^{\,4} & \bar{p}_4^{\,4} & \cdots & \bar{p}_{q-1}^{\,4} \\ \vdots & \vdots & \ddots & \vdots \\ \bar{p}_2^{\,q-1} & \bar{p}_4^{\,q-1} & \cdots & \bar{p}_{q-1}^{\,q-1} \end{bmatrix},\ \mathbf{p}^n = \begin{bmatrix} \bar{p}_2^{\,n} \\ \bar{p}_4^{\,n} \\ \vdots \\ \bar{p}_{q-1}^{\,n} \end{bmatrix}^{\mathrm{T}},\ \text{and}\ \mathbf{\Gamma} = \frac{1}{2}\begin{bmatrix} v_2^{-2}\,\Gamma((2+1)/2) \\ v_2^{-4}\,\Gamma((4+1)/2) \\ \vdots \\ v_2^{-(q-1)}\,\Gamma(((q-1)+1)/2) \end{bmatrix}$$

where the superscript T is used for a transpose and $\bar{p}_2 \equiv 1$. Then, Formula (9) is expressed by $\mathbf{A}\mathbf{w} = \mathbf{\Gamma}$,

$\bar{\mathbf{p}}^{q+1}\mathbf{w} = (2v_2^{q+1})^{-1}\Gamma(((q+1)+1)/2)$, and $\sum_i w_i = \sqrt{\pi}$. If we eliminate $\mathbf{w}$ from the first two relations, we obtain a relation possessing only the variable $v_2$ with parameters $\bar{p}_{2(i+1)}$,

$$\bar{\mathbf{p}}^{q+1}\mathbf{A}^{-1}\mathbf{\Gamma} = (2v_2^{q+1})^{-1}\Gamma(((q+1)+1)/2). \tag{10}$$

Once we find the solution $v_2$ from (10), we can obtain the weight coefficients from $\mathbf{w} = \mathbf{A}^{-1}\mathbf{\Gamma}$ and $w_1 = 1 - \sum_{k=2}^{q} w_k$; and the discrete velocities from (8).

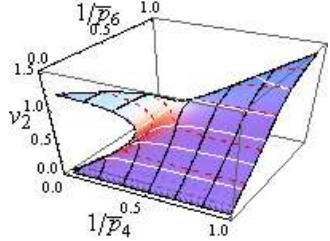

FIG. 1. (Color online) 3-dimensional plot generated by (10) when $q = 7$.

We plot Formula (10) for $M^1|7$ in FIG. 1. We will compare a part of this figure with the solution of $M^1|5$. We have obtained the discrete velocities and the weight coefficients of one-dimensional models. Therefore, we can easily construct higher-dimensional models from the one-dimensional ones according to tensor products of one-dimensional velocities [10].

From now on, we find $f^{eq}_E$ of Formula (2) to prepare discretized equilibrium distributions. For any multi-dimensional variable $\mathbf{x}$, the $N$-th order Taylor expansion (TE) of a function $g$ about $\mathbf{x} = \mathbf{x}_0$ is

$$g^{(N)}(\mathbf{x}) = \sum_{n=0}^{N} \frac{1}{n!} \left[ \left( \mathbf{x} \cdot \frac{\partial}{\partial \mathbf{x}'} \right)^n g(\mathbf{x}') \right]_{\mathbf{x}'=\mathbf{x}_0}. \tag{11}$$

Let $\mathbf{y}$ be a $D+1$ dimensional variable such that $\mathbf{y} = (\mathbf{u}, \theta)$ by combining the $D$-dimensional macroscopic velocity $\mathbf{u}$ and the temperature $\theta$. Let $f^{eq}_{TE}{}^{(N)}$ be the $N$-th order TE of the MB distribution about $\mathbf{y}_0 = (\mathbf{0}, \theta_0)$. Then, we obtain $f^{eq}_{TE}{}^{(N)}(\mathbf{y}) = g^{(N)}(\mathbf{y})$ by assuming that $\mathbf{u}$ and $\theta$ are infinitely small quantities of the same order. If we assume $\mathbf{u}$ and $\theta^{1/2}$ are the same order, we obtain another TE $f^{eq}_{HE}{}^{(N)}(\mathbf{z}) = g^{(N)}(\mathbf{z})$ about $\mathbf{z}_0 = (\mathbf{0},0)$ by defining $\mathbf{z} = (\mathbf{u}, \sigma)$ where $\varepsilon\sigma^2 = \theta - 1$ and $\varepsilon = \pm 1$. We emphasize the latter expansion is identical to the Hermite expansion (HE) of order $N$. The expansions $f^{eq}_{TE}{}^{(N)}$ and $f^{eq}_{HE}{}^{(N)}$ satisfy

$$\int \mathbf{v}^m f^{eq}(\mathbf{v}) d\mathbf{v} = \int \mathbf{v}^m f^{eq}_E(\mathbf{v}) d\mathbf{v} \tag{12}$$

for $0 \leq m \leq N$. This is clear if we regard $f^{eq}(\mathbf{v})$ as an infinite series expansion in $\mathbf{u}$ and $\theta$. The evaluated result of the left hand side of (12) contains the terms $\mathbf{u}^\alpha \theta^{[\beta]}$ such that $\alpha + 2\beta = m$ for $\alpha$ and $\beta$ are non-negative integers. Therefore, if the terms $\mathbf{u}^\alpha \theta^{[\beta]}$ are included in the series expansions, Formula (12) is satisfied. Note that $f^{eq}_{HE}{}^{(N)}$ itself is a polynomial of degree $N$ of $\mathbf{v}$, however, $f^{eq}_{TE}{}^{(N)}(\mathbf{v})$ is of degree $2N$ appearing in $\partial^N f^{eq}/\partial \theta^N |_{\theta=\theta_0}$.

Consequently, if we use $\mathrm{M}^1|q$ obtained by (9) with $f^{eq}_{TE}{}^{(N)}$, it is guaranteed that the moments evaluated from $f^{eq}_i$ are identical to those from $f^{eq}$ up to the $m$-th order where $m \leq \min[N, q+2-2N]$ and $\min[x,y]$ gives the smallest element between $x$ and $y$. The reason is that $f^{eq}_{TE}{}^{(N)}$ satisfies (12) for $m \leq N$ and $\mathrm{M}^1|q$ satisfies $\Xi(q+2)=0$ under the condition that $f^{eq}_{TE}{}^{(N)}$ itself is a polynomial of degree $2N$. Similarly, if we use $f^{eq}_{HE}{}^{(N)}$, we have the condition of satisfaction, $m \leq \min[N, q+2-N]$. Therefore, we can obtain sufficiently accurate models of the thermal lattice Boltzmann equation by controlling $q$ and $N$. The use of $f^{eq}_{HE}{}^{(N)}$ between TE and HE is optimal to reduce the number of discrete velocities in a given level of accuracy. However, from the viewpoint of stability, $f^{eq}_{TE}{}^{(N)}$ performs better. We will demonstrate it with simulation results. It seems the higher order terms in $\theta$ appearing in $f^{eq}_{TE}{}^{(N)}$ stabilize the models of the LBE. Note that the discretized equilibrium distribution is $f^{eq}_i = \rho w_i P(v_i)$ where $P(v) = \theta^{1/2} \exp(v^2) f^{eq}_E(v)$ and $f^{eq}_E$ is $f^{eq}_{TE}$ or $f^{eq}_{HE}$.

We explicitly present some models. We will give only the values of $v_2$ and $\bar{w}_{2i} (\equiv w_{2i}/\sqrt{\pi})$ because the others are easy to obtain by (7) and $\bar{w}_1 = 1 - \sum_{k=2}^{q} \bar{w}_k$. When $q=3$, we obtain the well-known solution of $v_2 = \sqrt{3/2}$ and $\bar{w}_2 = 1/6$. When $q=5$, we get

$$\begin{aligned}
v_2 &= (3+3r^2 \pm \chi)^{1/2}/2, \\
\bar{w}_2 &= [9r^4 - 27r^2 - 6 \mp (3r^2-2)\chi]/[300r^2(r^2-1)], \\
\bar{w}_4 &= [6r^4 + 27r^2 - 9 \mp (2r^2-3)\chi]/[300(r^2-1)]
\end{aligned} \quad (13)$$

where $\chi = (9r^4 - 42r^2 + 9)^{1/2}$ and $r = 1/\bar{p}_4$. The weight coefficients $\bar{w}_i$ are the same to those from the entropic method [10,11], however, instead of their fixed reference temperature for *isothermal* models, we provide the speeds of the discrete velocities for *thermal* models.

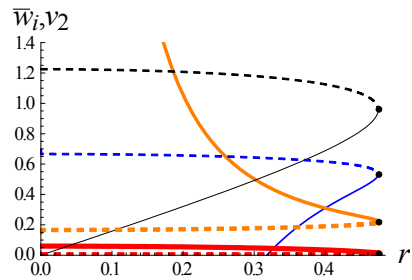

FIG. 2. (Color online) The graph of (13) is drawn. Two families are plotted and distinguished by dashed and solid lines. The thinner black line is for $v_2$, the thin blue for $\bar{w}_1$, the medium orange for $\bar{w}_2$, the thicker red for $\bar{w}_4$. The black dots indicate $\bar{w}_i$ and $v_2$ merged into identical values.

We define a ghost velocity by $v_i$ whose weight coefficient $\bar{w}_i$ is very small. For example, we can have a $\mathrm{M}^1|q$ having a pair of ghost velocities $v_{q-1}$ and $v_q$. When $\bar{w}_q$ approaches zero, the solution of (9) for a $\mathrm{M}^1|q$ satisfies $\Xi(q+1) = \Gamma(q+2/3)$ in addition to the system of equations for a $\mathrm{M}^1|(q-2)$. Consequently, the solution for a $\mathrm{M}^1|q$ having a pair of ghost velocities approaches the solution for a $\mathrm{M}^1|(q-2)$ with giving us additional information $\bar{w}_q$ and $v_q$. This exactly occurs in the $\mathrm{M}^1|5$ expressed by (13) drawn in FIG. 2. When $r$ approaches zero, the solution represented by dashed lines approaches that of the $\mathrm{M}^1|3$. And the solution of the $\mathrm{M}^1|7$ in FIG. 1 approaches that of the $\mathrm{M}^1|5$ when $1/p_6$ approaches zero. Therefore, it is better to use a solution set without ghost velocities to avoid the downgrade of the number of discrete velocities. We will show the stability issue related to the ghost velocities.

A 1-dimensional shock tube having a geometry of linear 1000 nodes was simulated by the two solutions of the $\mathrm{M}^1|5$ with $r=1/3$ and $f^{eq}_{HE}{}^{(3)}$. The initial condition is a density step such that $C_L = \{\rho = p = \bar{\rho}, \theta = 1, u = 0\}$ with $\bar{\rho} = 3$ for $X < 500$ and $C_R = \{\rho = p = \theta = 1, u = 0\}$ for $X \geq 500$. The boundary condition is $C_L$ with $\bar{\rho} = 3$ at $X = 1$ and $C_R$ at $X = 1000$. The relaxation time is $\tau = 1$. We observe the fluctuation only in the case of the solution having ghost velocities in FIG. 3. Note that the two solutions are both unstable if we increase $\bar{\rho}$ even to 4.

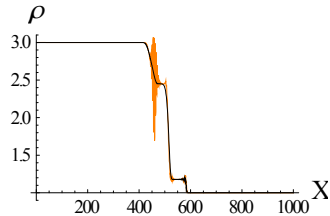

FIG. 3. (Color online) Simulation of shock tube obtained by $\mathrm{M}^1|5$. The black and orange lines are the results from the solutions of the dashed and the solid lines of FIG. 2, respectively.

The same shock tube problem was simulated again by $\mathrm{M}^1|5$, $\mathrm{M}^1|7$, and $\mathrm{M}^1|11$ with the ratios of discrete velocities and the base discrete velocity $(\bar{p}_4, v_2) = (3, 0.553432)$, $(\bar{p}_4, \bar{p}_6, v_2) = (2, 3, 0.846393)$, and $(\bar{p}_4, \bar{p}_6, \bar{p}_8, \bar{p}_{10}, v_2) = (2, 3, 4, 5, 0.685900)$, respectively, and compared with the analytical solution of the Riemann problem. Note that we present only the values of $v_2$ and $\bar{p}_{2i}$ to describe a specific model because of the limit of space. We can easily calculate the weight coefficients by $\mathbf{w} = \mathbf{A}^{-1}\mathbf{\Gamma}$. The simulation results are in excellent agreement to the analytical solution of the Riemann problem except the result obtained by a model having the second-order moment accuracy. For the physical properties $\rho$, $p$, $\theta$, and $u$, the subscripts 1 and 2 will be used to indicate that the values are extracted from the positions $X_1 = 430$ and $X_2 = 650$, respectively, where the plateaus appear. When $(q, N, m_{max}) = (5, 3^*, 3)$, $(7,3,3)$, and $(11,4,4)$, the results are $\rho_1 = 2.46$, $\rho_2 = 1.18$, $p_{1,2} = 1.65$, $\theta_1 = 0.67$, $\theta_2 = 1.40$, and $u_{1,2} = 0.22$ as in the solution of the Riemann problem. When $(q, N, m_{max}) = (5, 2, 2)$, the results are $\rho_1 = 2.43$, $\rho_2 = 1.18$, $p_{1,2} = 1.64$, $\theta_1 = 0.68$, $\theta_2 = 1.39$, $u_1 = 0.23$, and $u_2 = 0.22$. Note that $N$ is the order of the TE ($N = 2, 3, 4$) and HE ($N = 3^*$); $m_{max}$ is the maximum order of the satisfied moment.

If we pass a critical value of the density ratio between the left and the right domains in the initial and boundary conditions of the shock tube problem, the models of the LBE become unstable. Also, the decrease of the viscosity below a critical value by decreasing the relaxation time $\tau$ makes the models unstable. By virtue of (10), we could find the models having high number of discrete velocities to obtain robustness. Moreover, we realized that $f^{eq}_{HE}{}^{(N)}$ is not optimal from the viewpoint of stability. The following simulation shows the robustness of our $M^1|21$ having $(\bar{p}_4, \bar{p}_6, \bar{p}_8, \bar{p}_{10}, \bar{p}_{12}, \bar{p}_{14}, \bar{p}_{16}, \bar{p}_{18}, \bar{p}_{20}, v_2) = (2, 3, 4, 5, 6, 7, 8, 9, 11, 0.372889)$ and our discretized equilibrium distribution $f^{eq}_{TE}{}^{(5)}$. Under the initial condition $C_L$ with $\bar{\rho} = 11$ for $X \geq 500$ and $C_R$ for $X \geq 500$; and the boundary condition $C_L$ with $\bar{\rho} = 11$ at $X = 1$ and $C_R$ at $X = 1000$, the simulation was stable with $\tau = 1$ as in FIG. 4. However, $f^{eq}_{HE}{}^{(N)}$ with $N \leq 10$ could not pass the simulation test with the same conditions. Note that $f^{eq}_{HE}{}^{(10)}$ includes all terms appearing in $f^{eq}_{TE}{}^{(5)}$. We emphasize the $M^1|5$ with $(\bar{p}_4, v_2) = (3, 0.553432)$ was unstable even for $\bar{\rho} = 4$.

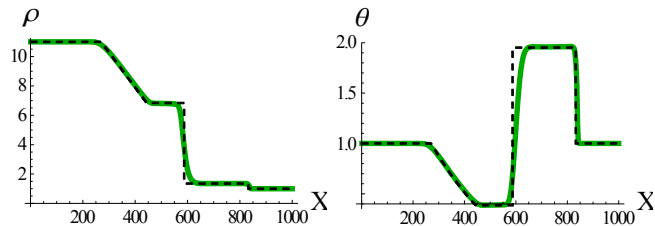

FIG. 4. (Color online) Analytical (black dashed line) and simulation (green solid line) results of shock tube.

In conclusion, we have derived a univariate polynomial equation providing models of the thermal LBE. Finding a model of any required level of accuracy, applicable to regular lattices, is reduced to only solving the univariate polynomial equation. This opens a way to construct robust models having high number of discrete velocities and applicable to regular lattices. We have presented discretized equilibrium distributions which are more robust than those obtained from the HE.